\documentclass[aps,amsfonts,pra]{revtex4-2}
\usepackage[utf8]{inputenc}
\usepackage{dsfont}
\usepackage{graphicx}
\usepackage{caption}
\usepackage{subcaption}
\newtheorem{algorithm}{Algorithm}
\begin{document}
\newcommand{\bra}[1]{\left\langle#1\right|}
\newcommand{\ket}[1]{\left|#1\right\rangle}
\newcommand{\abs}[1]{\left|#1\right|}
\newcommand{\mean}[1]{\left\langle #1\right\rangle}
\newcommand{\braket}[2]{\left\langle{#1}|{#2}\right\rangle} 
\newcommand{\commt}[2]{\left[{#1},{#2}\right]}
\newcommand{\tr}[1]{\mbox{Tr}{#1}}
\title{Two-Qutrit entanglement: 56-years old algorithm challenges machine learning}
\author{Marcin Wie\'sniak}\affiliation{Institute of Theoretical Physics, Faculty of Mathematics, Physics and Infortmatics, University of Gda\'nsk, 80-308 Gda\'nsk, Poland}
\affiliation{International Centre for Theory of Quantum Technologies, University of Gda\'nsk, ul. Ba\.zy\,nskiego 1A, 80-309 Gda\'nsk, Poland}
\date{\today}
\begin{abstract}
Classifying states as entangled or separable is a highly challenging task, while it is also one of the foundations of quantum information processing theory. This task is higly nontrivial even for relatively simple cases, such as two-qutrit Bell-diagonal states, i.e., mixture of nine mutually orthogonal maximally entangled states. In this article we apply the Gilbert algorithm to revise previously obtained results for this class. In particular we use ``cartography of entanglement'' to argue that most states left in  [Hiesmayr, B. C. {\em Scientific Reports} {\bf 11}, 19739 (2021)] as unknown to be entangled or separable are most likely indeed separable, or very weakly entangled. The presented technique can find endless applications in more general cases.
\end{abstract}

\maketitle

\section{Introduction}
Entanglement is probably the most striking feature of quantum mechanics. Initially, it was introduced for pure states, as notion that system are described only in reference to each other \cite{EPR}. This understanding was extremely convenient, as it was immediate to recognize an entangled state, and as it turned out, \cite{PureUnique}, in the bipartite case it can be quantified uniquely as the entropy of reduced states of subsystems. 

However, as the interest in foundations of quantum information theory grew, the issue got more complicated. For mixed states, it is not sufficient to study the deficit of information about individual subsystems at the expense of correlations. Instead, one needs to consider if the correlations can be reconstructed by means of local operations and classical communication (LOCC). A strong indicator of presence of entanglement in a state is a negative partial transpose (NPT) \cite{Peres,Horodecki}. Still, it turned out, except for dimensions of subsystems $2\times 2$ an $2\times 3$, entangled states with positive partial transpose (PPT) exist \cite{PPT}. This form of entanglement has been recognized as {\em bound} entanglement, which cannot be distilled to maximally entangled states (so-called Bell states) \cite{Distill}, as opposed to {\em free} NPT entanglement. Technically, remains an open question if bound entangled states with a negative partial transpose exists, but we will hereafter assume that all NPT entanglement is distillable.

More generally, all positive, but not completely positive maps were found to identify some form of entanglement \cite{Horodecki}. Alternatively, any entangled state yields a negative mean value of an entanglement witness operator, for which all separable (non-entangled states) give nonnegative means (Hereafter, we will adapt the opposite convention). The two methods happen to be related to each  other by the Jamio\l kowski-Choi isomorphism \cite{Jamiolkowski,Choi}. However, until recently there were no practical means to construct a witness tailored for a given state.

Further complications arose with attempts to give reasonable measures of entanglement. A list of requirements that any such measure must satisfy were formulated and two operational measures were defined: distillable entanglement \cite{DistillableEntanglement} and entanglement cost \cite{EntanglementCost}. The former quantifies the number of Bell states  required to prepare a copy of a given state in the LOCC, while the latter gives the number of them in the opposite process. Other measures were proposed, for example with the notion of a closest separable state (CSS) \cite{RelativeEntropy}, or as extensions of a unique measure for pure states with a convex roof extension \cite{ConvexRoof}. Still, the impossibility of considering all possible state manipulation schemes, decompositions into pure states and apparent computational difficulty in computing CSS with respect to some metrics prevented all of these propositions to become universal and practical.

Modern computing techniques could be  employed to the problem of separability in specific cases. For example, there was a number of attempts to train neural networks to distiguish entagled states from separable ones \cite{ML1,ML2,ML3,ML3AA,Hiesmayr,ML3A,ML4,ML5,ML6,ML7,ML8,ML9,ML10,ML11}. This approach poses its own set of challenges. First, as every use of machine learning (ML) techniques is application-specific, relatively little can be said about the optimality of the topology of the neural network, the form of the activation function, etc.. The other issue is that the result of ML is dependent on the training data, which can be very limited for the entanglement-vs-separability problem. A network can become overtrained with a fixation on the presented pattern. Finally, neural networks are not designed to estimate the missing continuous value of  function, which makes them less capable of providing the interpolation of the amount of entanglement, although there were some attempts of such tasks.

Recently, other strategies for classifying states as entangled or separable, based on variational algorithms \cite{Mirko} and semi-definite programming \cite{Otfried}, were proposed. 

In this articles we compare the algorithmic approach described in details below with ML results obtained by Hiesmayr \cite{Hiesmayr}, who focused on ``the magic simplex'', or Bell-diagonal states of three qubits. The task is to classify states as free entangled, bound entangled, or separable. Our methodology is decribed below, followed by a presentation of results and conclusions.
\section{Hilbert-Schmidt distance and Gilbert's algorithm}
The Hilbert-Schmidt distance is defined simply as
\begin{eqnarray}
D_{HS}(A,B)=&&\sqrt{\text{Tr}((A-B)(A-B)^\dagger)},\nonumber\\
=&&\sqrt{\sum_{i,j}|(A-B)_{i,j}|^2}.
\end{eqnarray}

This figure of merit has many advantages. Obviously, it is a well-defined distance, being just a variant of the usual Euclidean metric. It also allows to use intuitive terms from usual geometry, such as spheres, angles, or intervals. It is also the only measure, which can be computed without a computationally expensive subroutine of matrix diagonalization. 

Unsurprizingly, it was suggested as a core of entanglement measure by Witte and Trucks \cite{WitteTrucks}.
This entanglement quantifier shall be defined as
\begin{equation}
    D_{HS}(\rho)=\min_{\sigma\in SEP}D_{HS}(\rho,\sigma),
\end{equation}
where the mainimum is taken over all separable states. However, it was shown by Ozawa \cite{Ozawa} that it does not comply with the requirement of contractivity under an operation applied to both states. Other problems may occur when migrating a state to another Hilbert space by adding or tracing out an ancilla. Still, given the advantages of Hilbert-Schmidt measure, it can be seen as an insightful quantifier of nonclassical correlations.  

In 1966 Gilbert \cite{Gilbert} has introduced an algorithm to estimate the distance between a given point and a convex set. It was demonstrated to be useful to show the membership of a state in an LOCC equivalence \cite{Shang}, and to optimize Bell operators \cite{Vertesi}. Finally, Pandya, Sakarya, and Wie\'sniak \cite{PandyaSakaryaWiesniak} discussed a direct use of the algorithm to estimate the distance of a given quantum state to the set of separable states. The algorithm can be  outlined as follows.
\begin{algorithm} (bipartite case):
\label{Gilbert}
\\
{\em Input:} Test state $\rho_0$, initial separable state $\rho_1$.\\
{\em Output:} Approximation of CSS $\rho_1$, list of squared distances to subsequent CSS approximations $l$.
\begin{enumerate}
    {\item Take a random pure $\rho_2=\ket{\varphi_A}\ket{\varphi_B}\bra{\varphi_A}\bra{\varphi_B}$, that will be referred to as a trail state.}
    {\item If the the preselection criterion, $\text{Tr}(\rho_0-\rho_1)(\rho_2-\rho_1)>0$ is not met, go to step 1 or abort if the HALT condition is met.}
    {\item Maximize $\text{Tr}(\rho_0-\rho_1)(\rho_2-\rho_1)>0$ with local unitary transformation (run Algorithm 3)}
    {\item Update $\rho_1\leftarrow p \rho_1+(1-p)\rho_2$ for $p$ minimizing $D(\rho_0,p \rho_1+(1-p)\rho_2)$.}
    {\item Every 50 corrections append $D(\rho_0,\rho_1)$ to $l$.}
    {\item If the HALT condition is not met, go to step 1, otherwise quit.}
\end{enumerate}
\end{algorithm}
Pure random states are generated in the following subroutine \cite{ZyczkowskiSommers}
\begin{algorithm}:
\\
{\em Input:} dimension $d$.\\
{\em Output:} pure $d$-dimensional state randomly generated with respect to the Haar measure $\ket{\psi}$.
\begin{enumerate}
    {\item Draw $2d$ random numbers $r_1,r_2,...,r_{2d}$ from a normal distribution centered at 0.}
    {\item Construct $\ket{\psi}=\{c_i\}_{i=1}^d=\{r_{2i-1}+\iota r_{2i}\}_{i=1}^{d}$.}
    {\item Normalize $\ket{\psi}$ so that $\braket{\psi}{\psi}=1$.}
\end{enumerate}
\end{algorithm}

Features and performance of algorithm \ref{Gilbert} have been studied in Ref. \cite{PandyaSakaryaWiesniak}, with two major updates introduced in this work. One of them is the optimization in Step 3, which, depending on $\rho_0$, can speed up the convergence up to approximately 50 times, in terms of the required number of trails. In the code used for this article the optimization was performed with the following algorithm.  
\begin{algorithm}:
\\
{\em Input:} state $\rho_2$.\\
{\em Output:} optimized state $\rho_2$.\\
For $j=1$ to 1500:\\
do\\
\begin{enumerate}
    {\item draw a random qutrit state $\ket{\psi}$ uniformly with respect to Haar measure.}
    {\item With $j$ odd construct
    \begin{equation}
        U=(\mathds{1}+(e^{\iota\pi/100})\ket{\psi}\bra{\psi}))\otimes\mathds{1}.
    \end{equation}
    With $j$ even construct
    \begin{equation}
        U=\mathds{1}\otimes(\mathds{1}+(e^{\iota\pi/100})\ket{\psi}\bra{\psi})).
    \end{equation}
    }
    {\item If $\text{Tr}(\rho_0-\rho_1)(\rho_2-\rho_1)>\text{Tr}(\rho_0-\rho_1)(U\rho_2U^\dagger-\rho_1)$, replace $U\leftarrow U^\dagger$}
    {\item replace $\rho_2\leftarrow U\rho_2U^\dagger$ until $\text{Tr}(\rho_0-\rho_1)(\rho_2-\rho_1)>\text{Tr}(\rho_0-\rho_1)(U\rho_2U^\dagger-\rho_1)$}
\end{enumerate}
done.
\end{algorithm}

The other important update concerns the analysis of list $l$ and is described below.

The algorithm provides three pieces of information that will be used here to classify states as entangled or separable. The first is, rather trivially, the last squared distance $D^2_{\text{Last}}$ found within a fixed number of corrections. For more than 1000 corrections, it is believed to be a fairly good representation of the actual distance. In particular, there are {\em deep} separable states, which in few corrections can be reconstructed nearly down to numerical precision, an hence, they could be safely assumed to be separable. As the algorithm in principle cannot reach $D^2=0$, this indicator tends to overestimate entanglement for states close the entangle-separable boundary. Our approach emphasis on the quantitative, rather than qualitative description of entanglement.

The second indicator is the distance decay estimate, $D^2_{\text{Est}}$. This is the other improvement with respect to Ref. \cite{PandyaSakaryaWiesniak}. The current recommendation is to maximize the linear regression coefficient,
\begin{equation}
    R(x,y)=\frac{\mean{x\odot y}-\mean{x}\mean{y}}{\sqrt{(\mean{x\odot x}-\mean{x}^2)(\mean{y\odot y}-\mean{y}^2)}},
\end{equation}
where $\odot$ denotes the entry-wise product of two lists, between the list of correction numbers $c$ and $1/(l-a)$ (each element of $l$ is shifted by $a$ and then inversed), with rejecting the first one third of the entries (denoted by $\tilde{cdot}$. We thus have
\begin{equation}
    D_{\text{Est}}^2=\text{agrmax}_a R(\tilde{c},1/(l-a)),0\leq a\leq \max l.
\end{equation}

Although, unlike the other two indicators, $D^2_{\text{Est}}$ does not certify anything, but is merely an result of a statistical analysis, we shall find it to be quite informative. With $R(x,a x+b)=1$, $a>1$ and $b$ being real numbers, we typically get $R\approx 0.998$ for 10000 corrections.

One advantage of $D_{\text{Est}}$ over $D_{\text{Last}}$ is that the distance decay estimate can be found to be 0, suggesting no entanglement in the studied state.

The third and final figure of merit considered here here is the witness distance estimate $D_{\text{Wit}}$. As Ref. \cite{Krammer} points out, if $\rho_{0,\text{CSS}}$ is the actual closest separable state, operator $W_0=\rho_0-\rho_{0,\text{CSS}}$ attains positive values only for entangled states, thus being an entanglement witness. In reality, $\rho_1$ is only an approximation of $\rho_{0,\text{CSS}}$, displacing and tilting the hyperplane represented by the witness. As stated in Ref. \cite{QuantumReports}, we then need to consider
\begin{equation}
W=\rho_0-\rho_1-\max_{\ket{\phi_1},\ket{\phi_2}}\mathds{1}\bra{\phi_1}\bra{\phi_2}\rho_0-\rho_1\ket{\phi_1}\ket{\phi_2},    
\end{equation}
with the maximum taken over all product states. Then
\begin{equation}
D_{\text{Wit}}=\max\left(0,\frac{\text{Tr}W\rho_0}{\sqrt{\text{Tr}(\rho_0-\rho_1)^2}}\right).    
\end{equation}
Again, this quantity depends on the quality of the approximation $\rho_1$, and provides a bound, this time lower, on the actual distance. As the only one of the three, It actually certifies the presence of entanglement. However, it is more challenging computationally than it seems. To claim to have found the global maximum of trigonometric polynomial we need to have a high confidence level that we did not obtain a value for any local extreme. This is obtainable, for example, by multiple repetitions of the maximization with randomized initial conditions.  Having taken care of it, we shall assume that we have indeed found the absolute maximum.

\section{``Magic Simplex'' states}
Equipped with the above tools to analyse entanglement, we will revisit certain classes of two-qutrit Bell diagonal states, also known as ``Magic Simplex'' states. 

Let us consider MES of two qutrits,
\begin{equation}
    \ket{\psi_{00}}=\frac{1}{\sqrt{3}}\sum_{i=0}^2\ket{ii},
\end{equation}
and two Weyl operators, $X=\left(\begin{array}{ccc}0&1&0\\0&0&1\\1&0&0\end{array}\right)$ and $Z=\left(\begin{array}{ccc}1&0&0\\0&\alpha&0\\0&0&\alpha\end{array}\right)$ with $\alpha=e^{2\pi \iota/3}$. The Bell basis is given by
\begin{equation}
    \{\ket{\psi_{ij}}\}_{i,j=0}^2=\{\mathds{1}\otimes X^iZ^j\ket{\psi_{00}}\}_{i,j=0}^2.
\end{equation}
Then the Bell-diagonal states are given by 
\begin{eqnarray}
&&\rho=\sum_{i,j=0}^2p_{ij}\ket{\psi_{ij}}\bra{\psi_{ij}},\nonumber\\
&&\sum_{i,j=0}^2p_{ij}=1, p_{ij}\leq 0.
\end{eqnarray}

In particular, Hiesmayr focused on four families of states. Family $A$ was given by
\begin{eqnarray}
    \rho(\alpha,\beta,\gamma)=&&(1-\alpha-\beta-\gamma)\frac{\mathds{1}}{9}\nonumber\\
    +&&\alpha\ket{\psi_{00}}\bra{\psi_{00}}\nonumber\\
    +&&\beta\ket{\psi_{01}}\bra{\psi_{01}}\nonumber\\
    +&&\gamma\ket{\psi_{02}}\bra{\psi_{02}},
\end{eqnarray}
and it was considered for $\gamma=0$. The  other three families are a part of Family $B$, which includes all the states in form
\begin{eqnarray}
\rho(\alpha,\beta,\gamma,\delta)&&=(1-\alpha-\beta-\gamma-\delta)\frac{\mathds{1}}{9}\nonumber\\
+&&\alpha\ket{\psi_{00}}\bra{\psi_{00}}\nonumber\\
+&&\frac{\beta}{2}(\ket{\psi_{01}}\bra{\psi_{01}}+\ket{\psi_{02}}\bra{\psi_{02}})\nonumber\\
+&&\frac{\gamma}{3}\sum_{j=0}^2\ket{\psi_{1j}}\bra{\psi_{1j}}\nonumber\\
+&&\frac{\delta}{3}\sum_{j=0}^2\ket{\psi_{2j}}\bra{\psi_{2j}}.
\end{eqnarray}

This family was studied in three different cases. Family $B_1$ had 
$\gamma=-\frac{1}{\sqrt{3}}$ and $\delta=0$, Family $B_2$ --$\gamma=-0.83$ and $\delta=0$, while $B_3$ included $\alpha=\frac{5}{3}(\sqrt{3}-1)$ and $\beta=-\frac{1}{10}$.

The aim of this work is to recover, or possibly expand on results presented recently by Hiesmayr in Ref. \cite{Hiesmayr}. For selected families, she first checked the states for negative partial transposition identifying all the states containing free entanglement (FREE). For the rest, which could be either bound entangled (BOUND) or separable (SEP), she collected any other known criteria of entanglement, the. Another portion of states was recognized as bound entangled in a numerical search of respective entanglement witnesses, reportedly being the most computationally demanding part. Finally, neural networks were trained to classify remaining unknown (UNKNOWN) states as SEP  or BOUND. This, however, required preassigning a class to each state, to be  changed later by the neural network.

\section{Results}
In this work we utilize a technique of entanglement charts. For randomly scattered states within a given set, we calculate $D_{\text{Last}}$, $D_{\text{Est}}$, and $D_{\text{Wit}}$. The found values are then interpolated, and the resulting charts can be compared with plots in Ref. \cite{Hiesmayr}.  The charts contain evidence, if not solid proofs, for BOUND and SEP regions. 

\subsection{Family A}
We first focus on Family $A$, in which we studied 1485 PPT states, each with up to 4000 corrections (the algorithm HALTS at $D^2(\rho_0,\rho_1)<10^{-7}$). Figure 1 presents a 3D plot of $D_{EST}$. It presents an almost hollow structure. This is due to the renderer omitting the extreme values for clarity and the convergence being slower at the boundary between separable and entangled states. The only prominent maxima correspond to small sets bound entangled states confirmed in Refs. \cite{Ref36,Ref37}. No entanglement witnesses were found. It is thus reasonable to assume that for this family bound entanglement exists only in neighbourhoods found in Refs. \cite{Ref36,Ref37}. 
\subsubsection{Close-up}
We have thus performed a more detailed study of 160 Family $A$ PPT states with $\alpha\in(0.175,0.225)$,  $\beta\in(-0.075,-0.025)$, and $\gamma=0$. We started with 90000 corrections for each state, and extended the study for six states, which had $D_{\text{Est}}>0$. However, within  HALT condition of up to 220000 corrections or 2 billion trail states we have found only three states with $D^2_{\text{Wit}}>0$. We thus leave the issue of the remaining five states unresolved by our method. Nevertheless, we remark that the plot for $D_{\text{Est}}$ reproduces the shape of the set of bound entangled states presented in Figure 3 of Ref. \cite{Hiesmayr}. It should be also noted that for the eight states with $D_{\text{Est}}>0$ the regression coefficient was greater than $0.9999$ The results for $D_{\text{Last}}$, $D_{\text{Est}}$, $D_{\text{Wit}}$ are presented in Figure 2.
\begin{widetext}
\centering
\begin{figure}[!h]
    \includegraphics{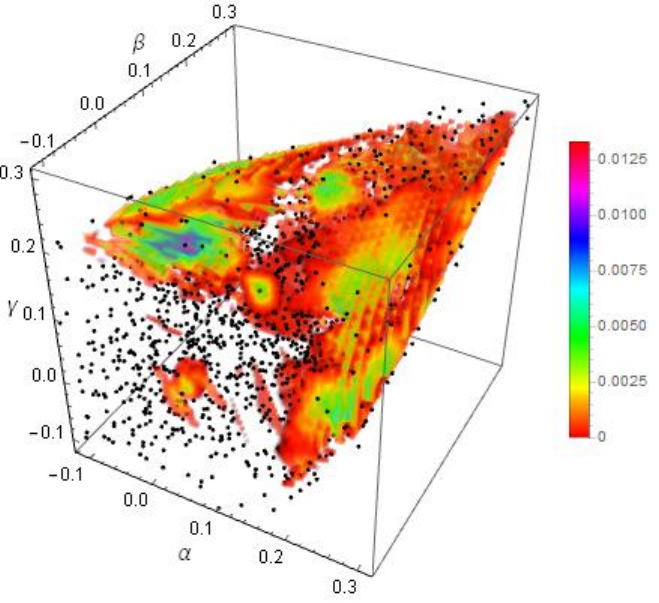}
    \caption{3D plot of $D_{\text{Est}}$ for PPT states from family A. Black dots represent studied states.}
    \label{Fig1}
\end{figure}

\begin{figure}[!h]
    \centering
    \begin{subfigure}[l]{0.3\textwidth}
    \includegraphics[width=60mm]{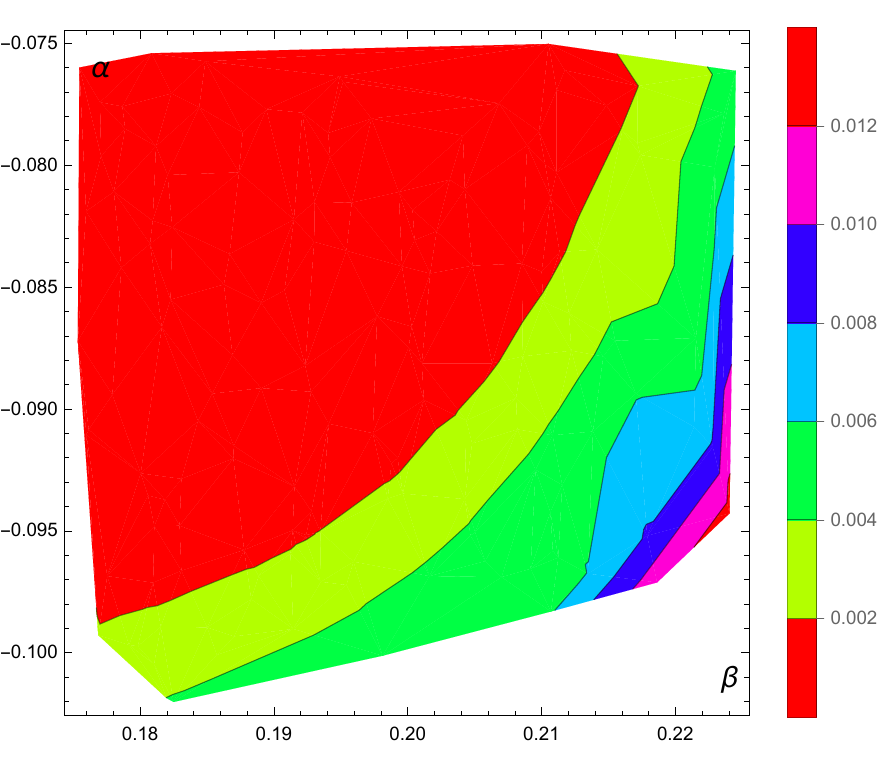}
    \end{subfigure}
    \hfill
    \begin{subfigure}[c]{0.3\textwidth}
    \includegraphics[width=60mm]{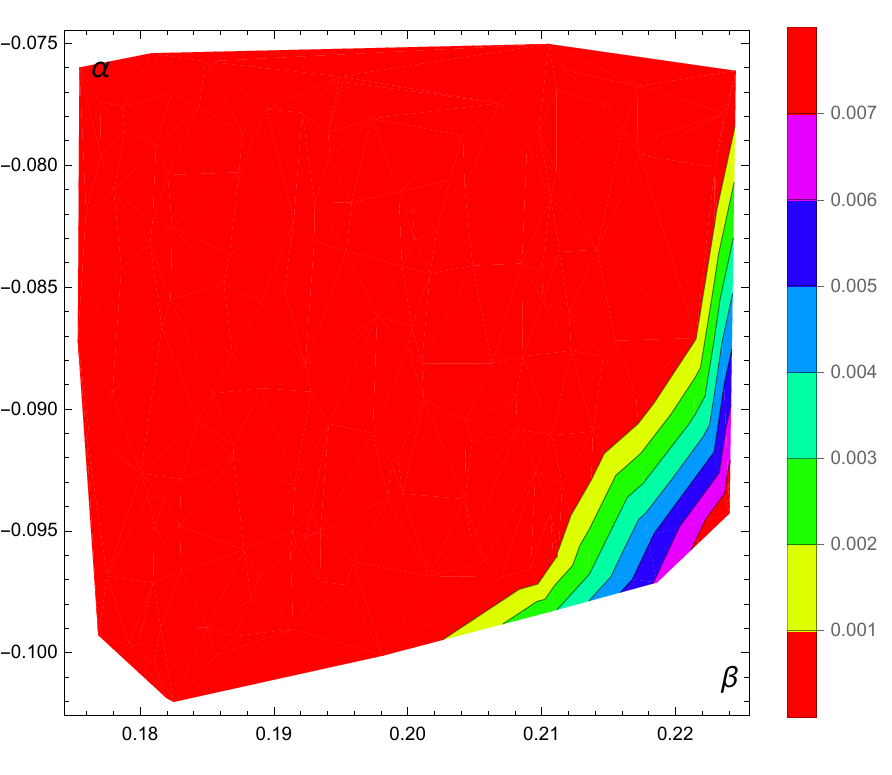}
    \end{subfigure}
    \hfill
    \begin{subfigure}[r]{0.3\textwidth}
    \includegraphics[width=60mm]{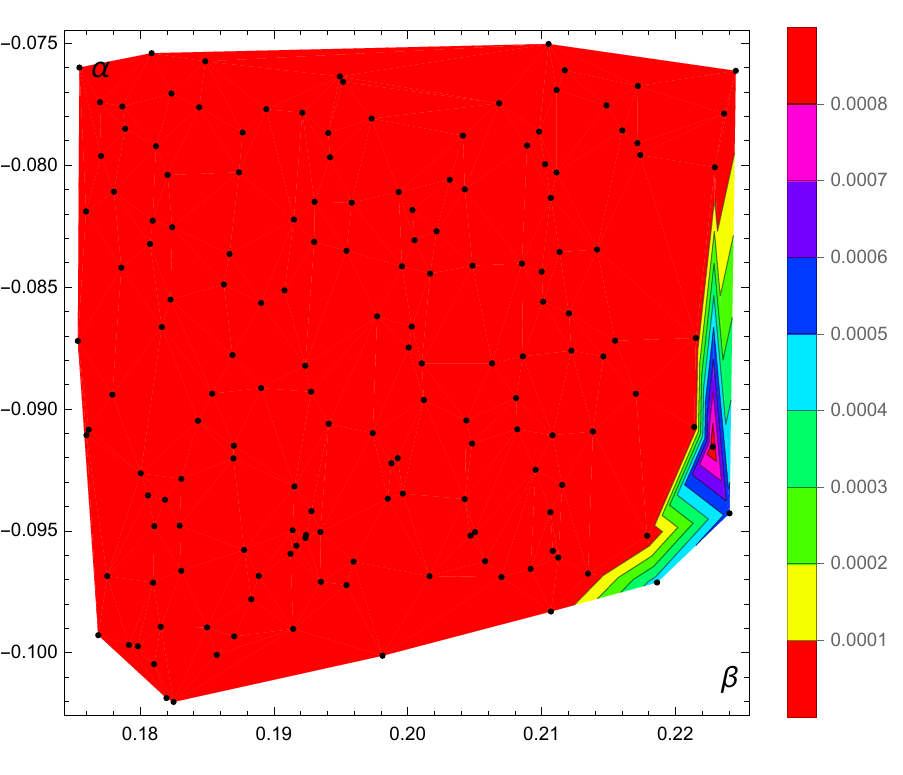} 
    \end{subfigure}
    \caption{The interpolated plots of $D_{\text{Last}}$ (left), $D_{\text{Last}}$ (center), and $D_{\text{Wit}}$ (right) for Family $A$ with $\gamma=0$ (only a region with PPT entangled state). Black dots represent studied states.}
    \label{Fig2}
\end{figure}
\end{widetext}
\subsection{Family $B$}
Subsequently, we have conducted the analysis for the remaining three families. Within each family we generated 300 states. For Family $B_1$, $B_2$, and $B_3$ we conducted up to 10000, 8000 and 10000 corrections respectively. The results are presented in Figure 3 and they seem to adequately reproduce the plots in Ref. \cite{Hiesmayr}. Optimization for states belonging to Family $B_2$ was significantly slower, meaning that they require significantly more trail states for a correction. Hence, the algorithm was HALTed with fewer corrections.
\begin{widetext}

\begin{figure}[!h]
    \begin{subfigure}[l]{0.3\textwidth}
    \includegraphics[width=60mm]{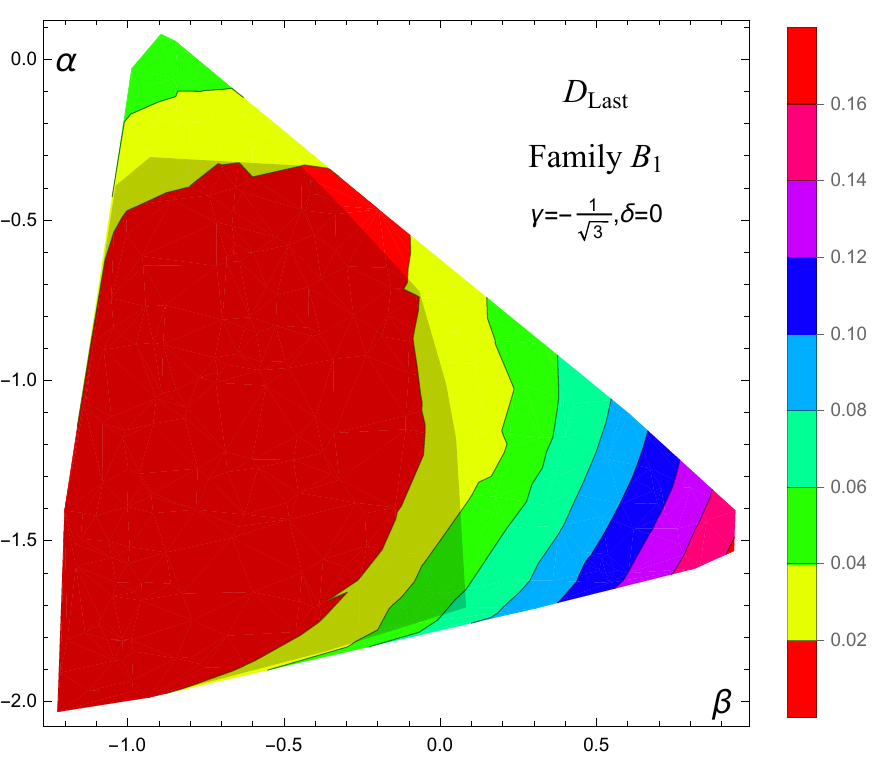}
    \end{subfigure}
    \hfill
    \begin{subfigure}[c]{0.3\textwidth}
    \includegraphics[width=60mm]{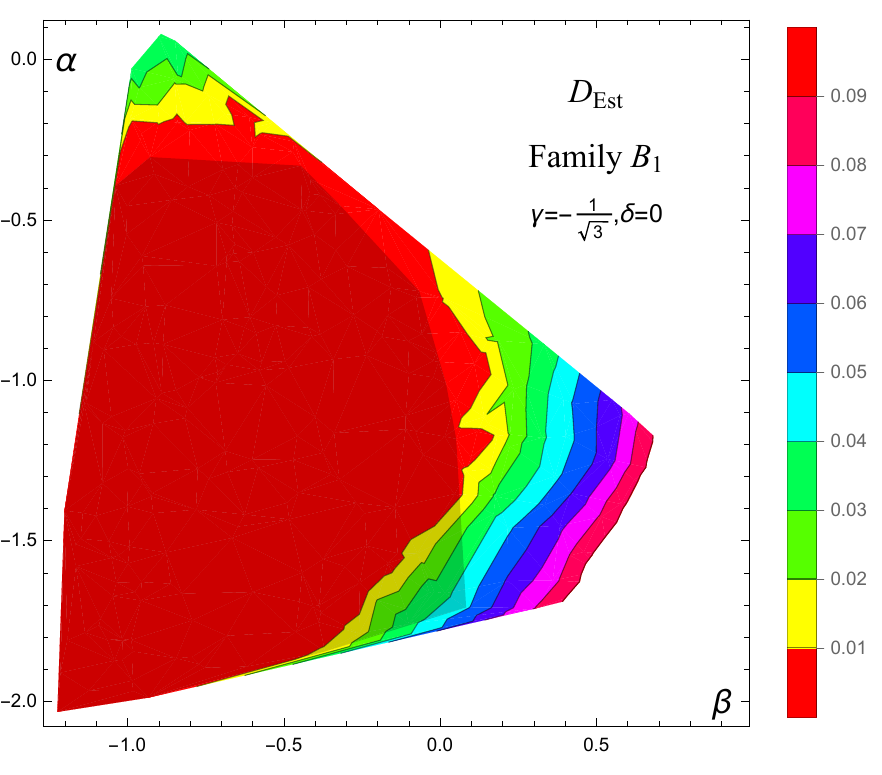}
    \end{subfigure}
    \hfill
    \begin{subfigure}[r]{0.3\textwidth}
    \includegraphics[width=60mm]{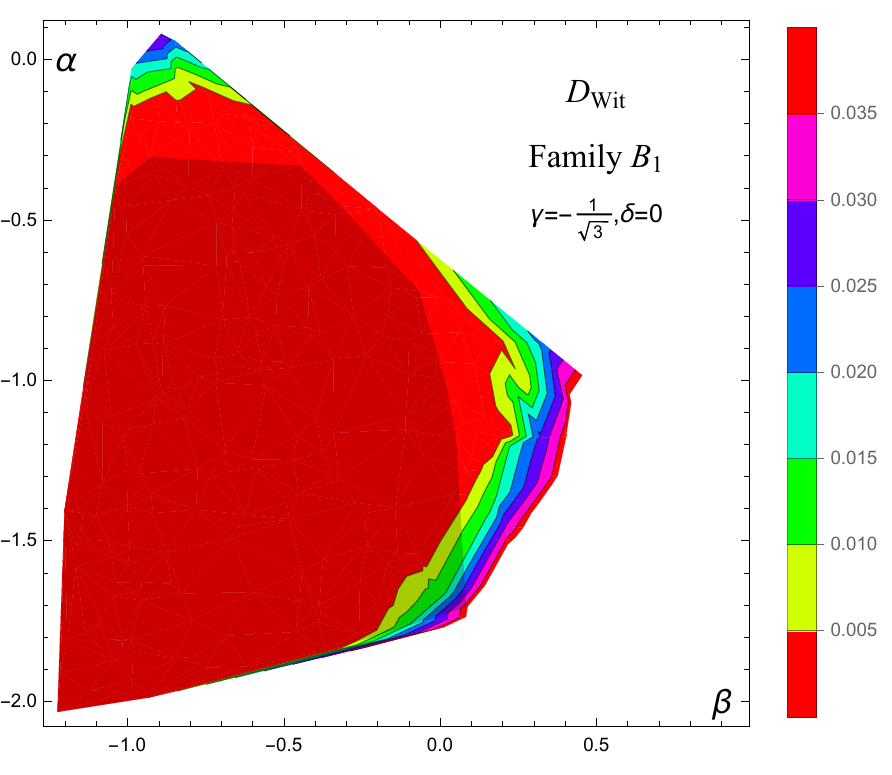}
    \end{subfigure}
    \vfill
    \begin{subfigure}[l]{0.3\textwidth}
    \includegraphics[width=60mm]{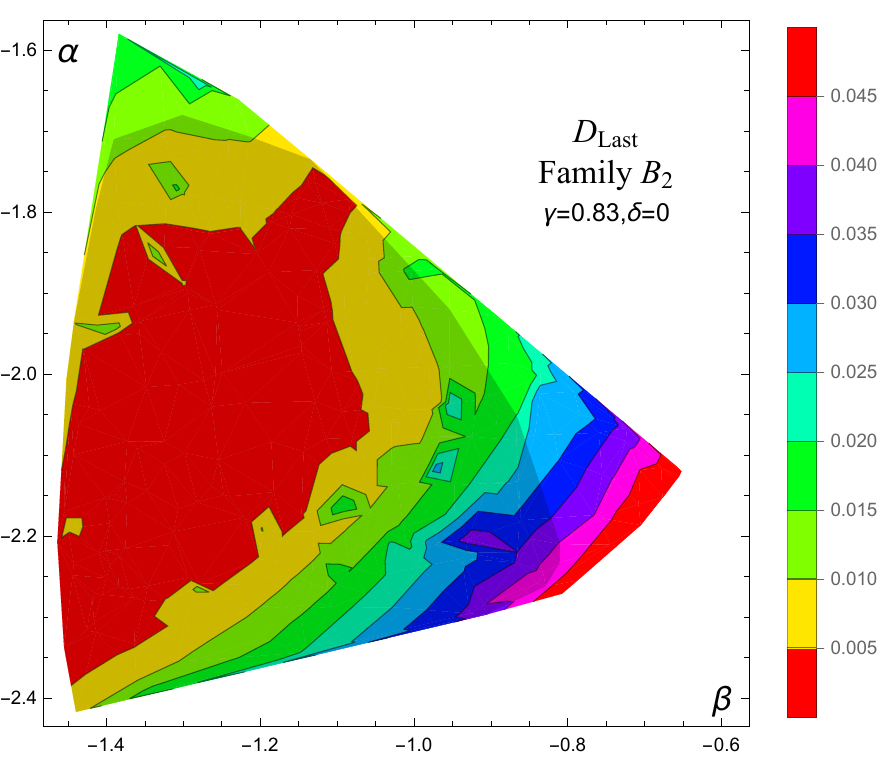}
    \end{subfigure}
    \hfill
    \begin{subfigure}[c]{0.3\textwidth}
    \includegraphics[width=60mm]{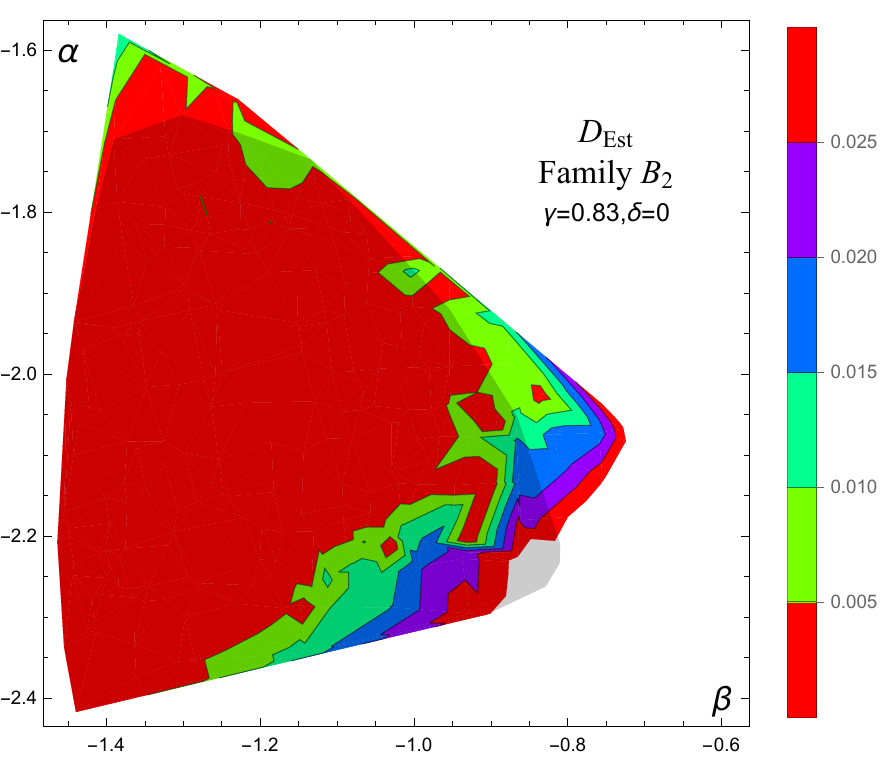}
    \end{subfigure}
    \hfill
    \begin{subfigure}[r]{0.3\textwidth}
    \includegraphics[width=60mm]{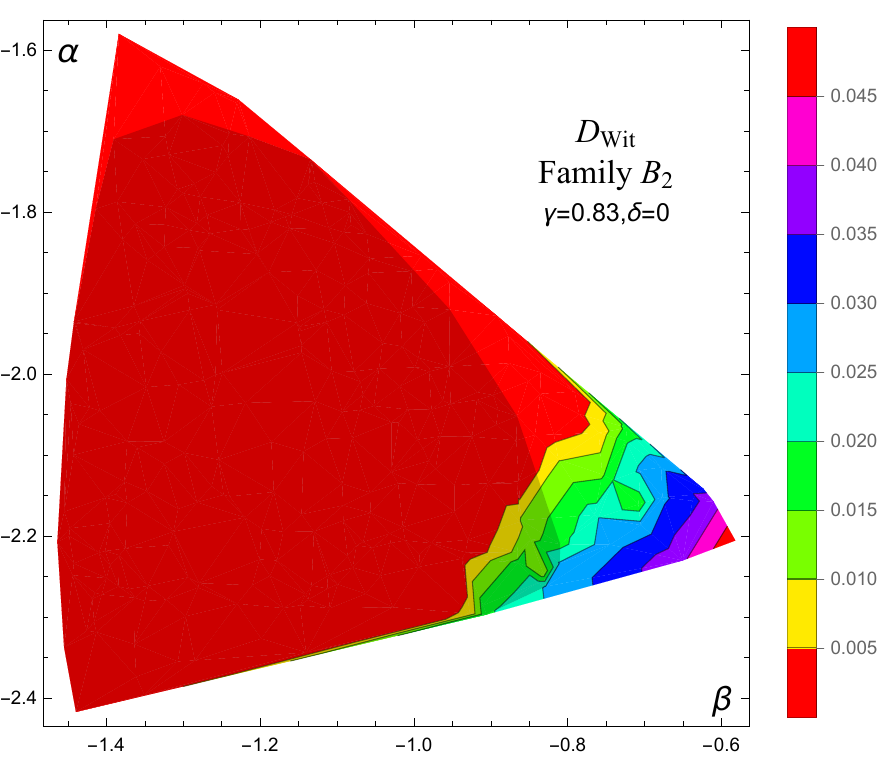}
    \end{subfigure}
    \vfill
    \begin{subfigure}[l]{0.3\textwidth}
    \includegraphics[width=60mm]{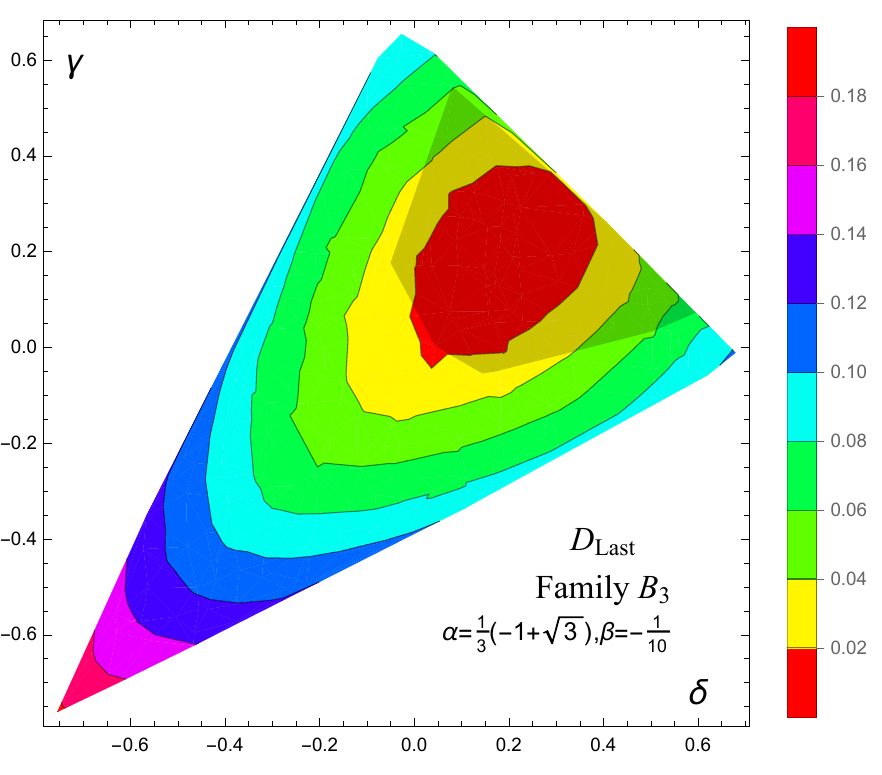}
    \end{subfigure}
    \hfill
    \begin{subfigure}[c]{0.3\textwidth}
    \includegraphics[width=60mm]{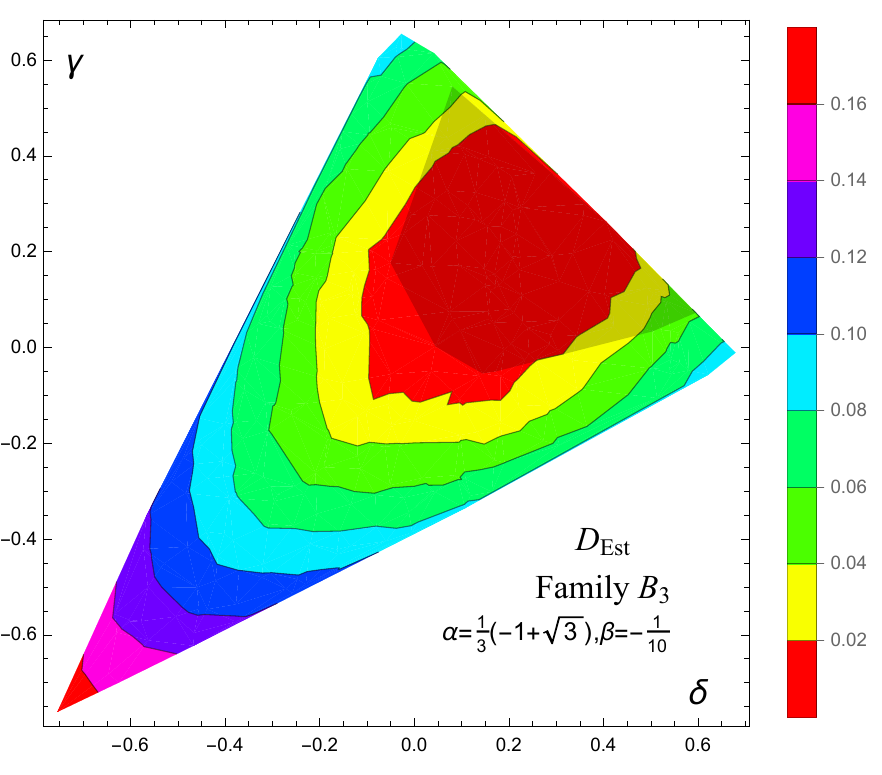}
    \end{subfigure}
    \hfill
    \begin{subfigure}[r]{0.3\textwidth}
    \includegraphics[width=60mm]{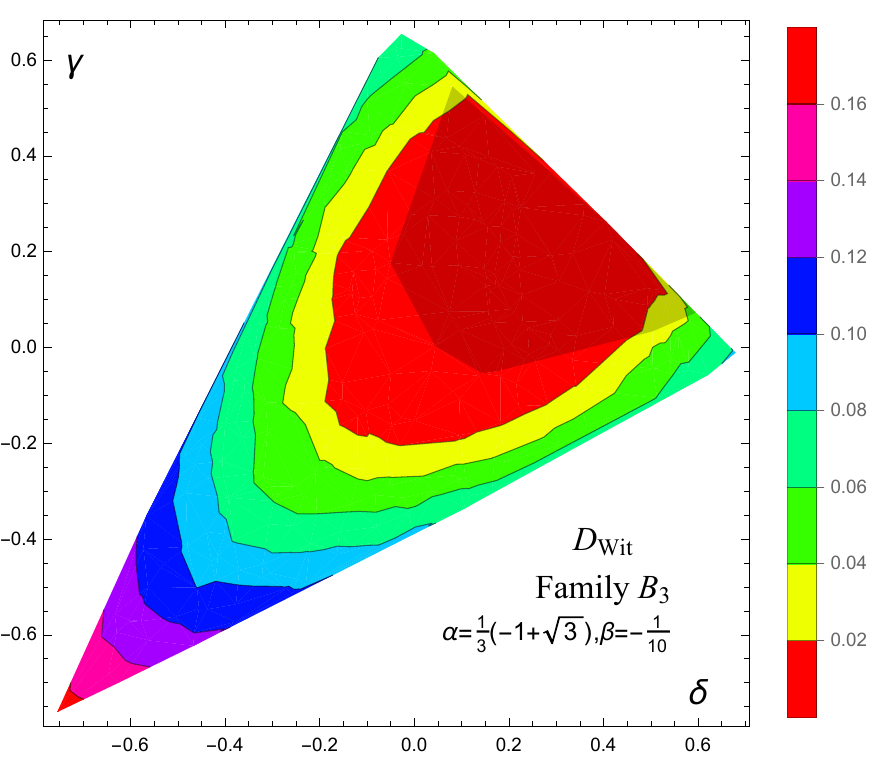}
    \end{subfigure}
    \caption{The interpolated plots of $D_{\text{Last}}$ (left), $D_{\text{Last}}$ (center), and $D_{\text{Wit}}$ (right) for Families $B_1$ (top), $B_2$ (middle) and $B_3$ (bottom) Shaded areas represent PPT states.}
    \label{Fig3}
\end{figure}
\end{widetext}

\subsubsection{Close-up of Family $B_3$}
However, we have performed an extensive study of the PPT part of Family $B_3$. This Family was chosen primarily because only for families $B_2$ and $B_3$ Hiesmayr presented machine learning results with two different preassigment strategies, ``random forest'' and ``nearest neighbours''. For Family $B_2$, one of the results suggests all states to be entangled, and both results contradict our findings. For Family $B_3$ the two approaches gave rather inconsistent results, which are aggregated in Figure 4. We have generated 160 states, which were studied up to 60000 corrections each. This number of corrections provided very high consistency between $D_{\text{Last}}$, $D_{\text{Est}}$, and $D_{\text{Wit}}$. We found 32 entanglement witnesses. The results are presented Figure 5. Importantly, the set of states recognized as entangled exceed the set bounded by witnesses presented in Ref. \cite{Hiesmayr}. In this case, the algorithm found a new set of bound states. For example, a state with coordinates $(\gamma,\delta)=(-0.0226,0.3067)$ was confirmed to be entangled while it was not detected in the original Reference. Also, he boundary between bound entangled and separable states seems to be curved, rather than straight. Given the geometric interpretation of the problem, this is actually expected.
\begin{figure}[!t]
    \centering
    \includegraphics[width=.5\textwidth]{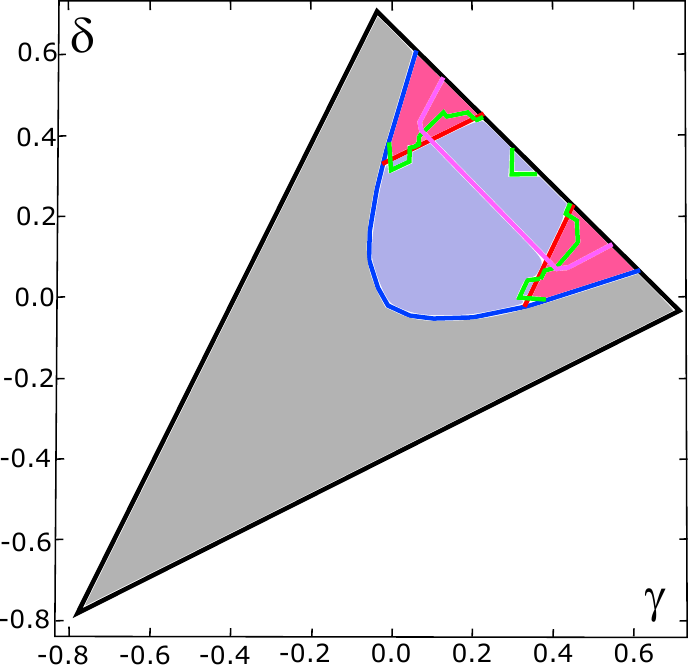}
    \caption{Clasification of Family $B_3$ according to Ref. \cite{Hiesmayr}. The color coding is as follows: gray: free entangled states, red: bound entangled states confirmed by linear witnesses, green and pink lines: boundaries of the set of PPT entangled states as suggested by machine learning, with ``nearest neighbours'' and ``random forresting'', respectively.}
    \label{BeatrixPlot}
\end{figure}
\begin{widetext}

\begin{figure}[!h]
\centering
    \begin{subfigure}[l]{0.3\textwidth}
    \includegraphics[width=60mm]{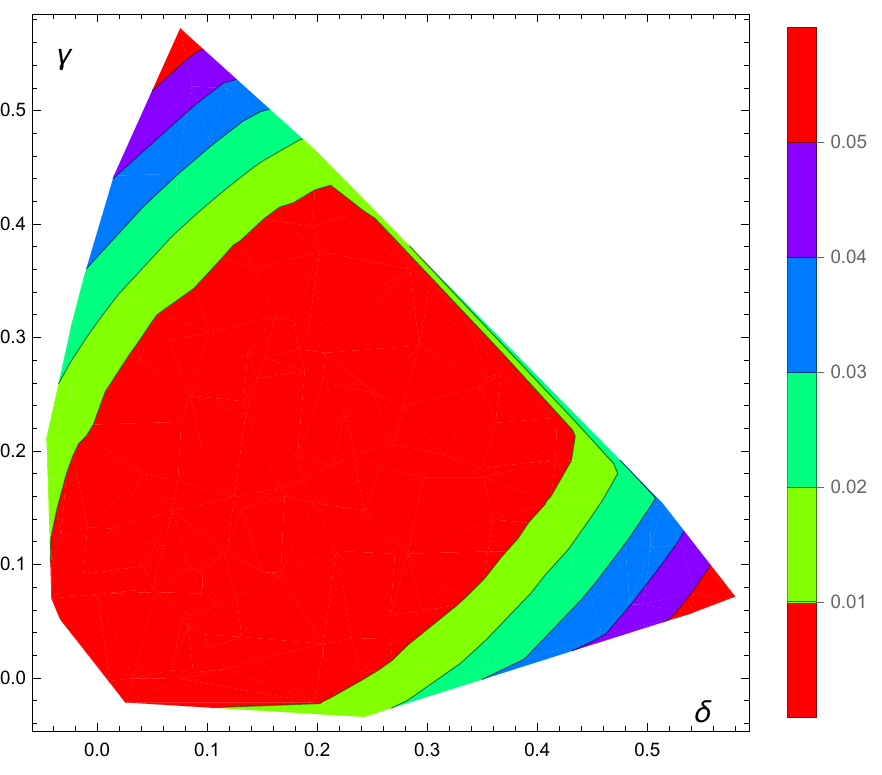}
    \end{subfigure}
    \hfill
    \begin{subfigure}[c]{0.3\textwidth}
    \includegraphics[width=60mm]{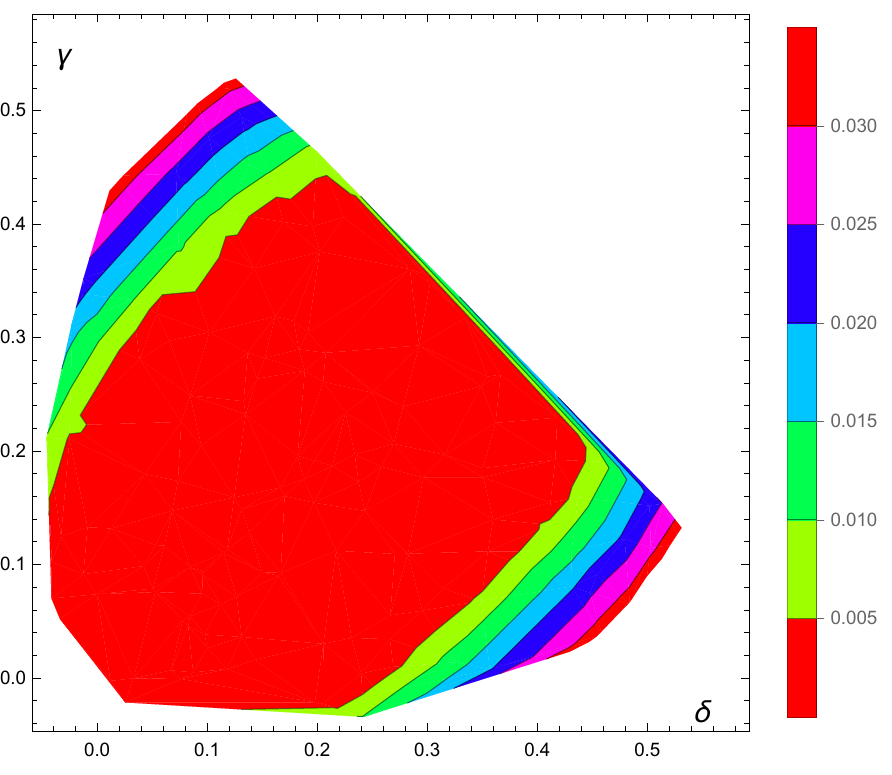}
    \end{subfigure}
    \hfill
    \begin{subfigure}[r]{0.3\textwidth}
    \includegraphics[width=60mm]{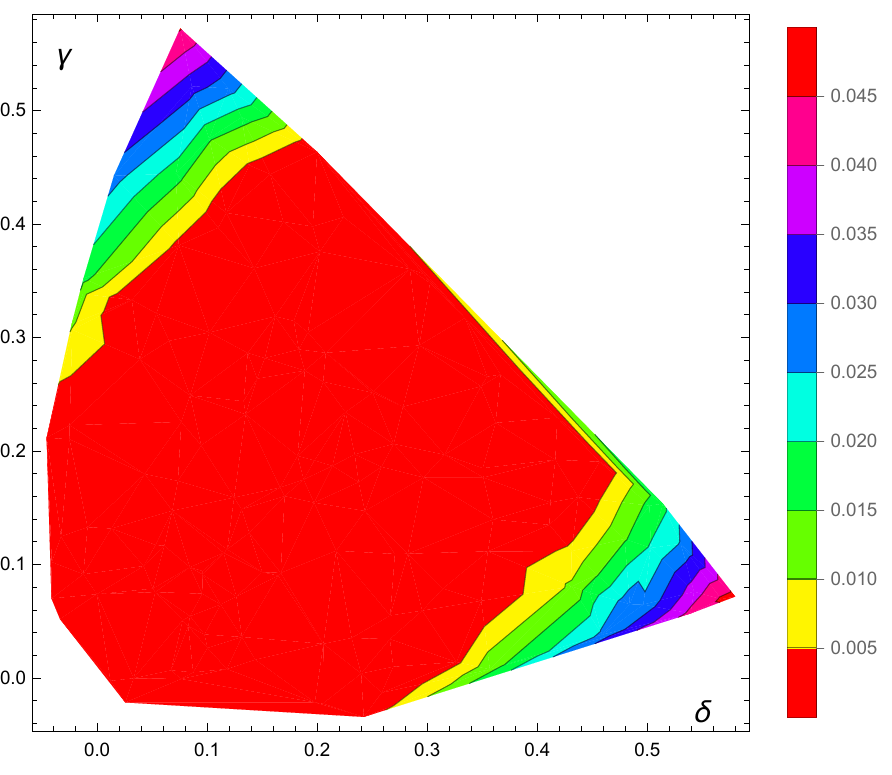}
    \end{subfigure}
    \caption{Interpolated plots of $D_{\text{Last}}$ (left), $D_{\text{Est}}$ (center), and $D_{\text{Wit}}$ (right) for PPT states of Family $B_3$.}
\end{figure}
\end{widetext}
\subsubsection{Finer study of Family $B_2$}
We also selected 34 states from Family $B_2$ that are close to to an expected boundary between separable and entangled states. We carried the calculation up to 40000 corrections. The results eliminated much of chaos visible in the middle row of Figure 3, but still the region of $D_{\text{Wit}}>0$ is smaller than as presented in Ref. \cite{Hiesmayr} (see Figure 6).
\subsection{Volumetry of bound entanglement in simplex}
In each case Hiesmayr gave a detailed estimate of the relative volume of bound entangled states. In this paper we rather avoid it  due to substantial differences in methodology. We do not perform a massive survey of states, but take a small sample and interpolate the results. Additionally, while the algorithm is capable of precisely estimating the distance for highly entangled states and deeply separable states, it leaves many boundary cases undecided. Moreover, we use three different estimators, each dependent on how far we go in the endless sequence of corrections. All of it thus draws the relative volumes somewhat arbitrary. The only case where we decided investigate this issue was  the full simplex parametrized by 
\begin{eqnarray}
\rho=&&\sum_{(i,j)\neq (2,2)}p_{ij}\ket{\psi_{ij}}\bra{\psi_{ij}}\nonumber\\
+&&\left(1-\sum_{(i,j)\neq (2,2)}p_{ij}\right)\ket{\psi_{22}}\bra{\psi_{22}}.
\end{eqnarray}
With $\{p_{ij}\}_{(i,j)\neq(2,2)}$ drawn at random uniformly, we generated 1000 PPT states, which took the total of 2456 physical states. This gives the ratio of .407 of states being PPT We run the algorithm up to 30000 corrections for each state and found that $D_{\text{Est}}$ was larger than 0 for 137 states, whereas $D_{\text{Wit}}>0$ was found in 45 cases. We also checked the parametrization including the white noise,
\begin{eqnarray}
\rho=&&\sum_{i,j=0}^2a_{ij}\ket{\psi_{ij}}\bra{\psi_{ij}}\nonumber\\
+&&\left(1-\sum_{i,j=0}^2a_{ij}\right)\frac{\mathds{1}}{9},
\end{eqnarray}
where each $a_{ij}\in\{-1/8,1\}$ was drawn randomly with a uniform distribution. Generating 1000 PPT states took 2398 physical states, which translates ito 0.417 states being PPT. Again, we have reconstructed CSSs up to 30000 corrections. As a result, we found 139 cases of $D_{\text{Est}}>0$ and $45$ cases of $D_{\text{Wit}}>0$.

\subsection{Dynamics of $D_{\text{Est}}$}
We also present the dynamics of $D_{\text{Est}}$ in function of the number of corrections. Figure 7 depitcs this quantity for three arbitrarily chosen states from Family $B$ and three from the cloeup of Family $A$ (Figure 2). The only criterion was that $D_{\text{Est}}$ is larger than 0. While it is difficult to characterize this behavior in general, one can conclude that for many states $D_{\text{Est}}$ stabilizes after sufficiently many corrections. It is thee case of 5 out of presented states. This again highligths the relevance of this quantity as an entanglement estimator.
\begin{widetext}
\begin{figure}[!h]
\centering
    \begin{subfigure}[l]{0.3\textwidth}
    \includegraphics[width=60mm]{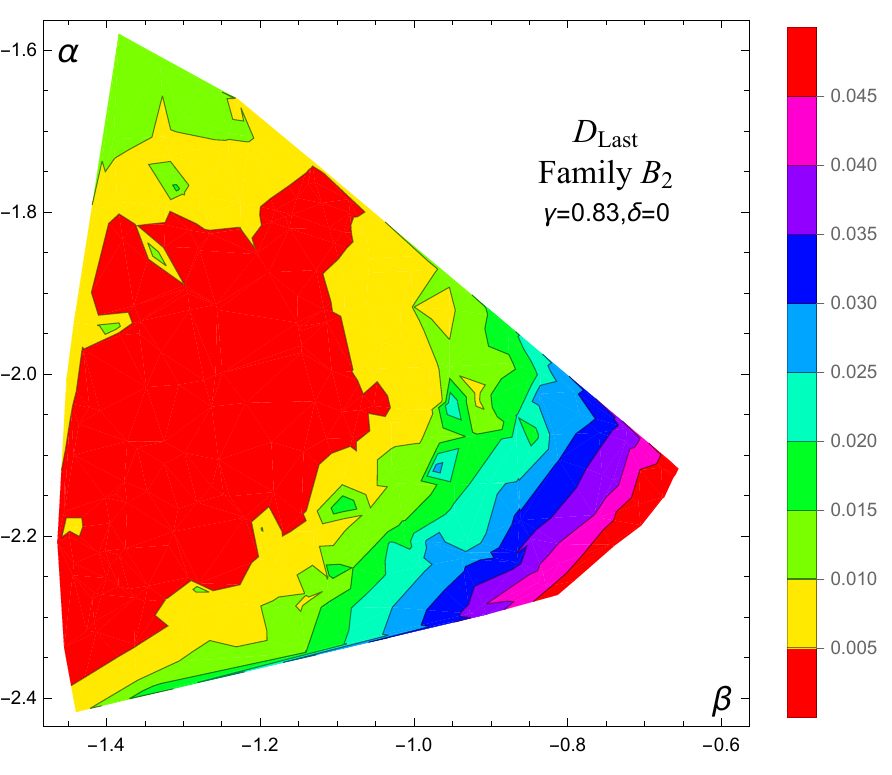}
    \end{subfigure}
    \begin{subfigure}[c]{0.3\textwidth}
    \includegraphics[width=60mm]{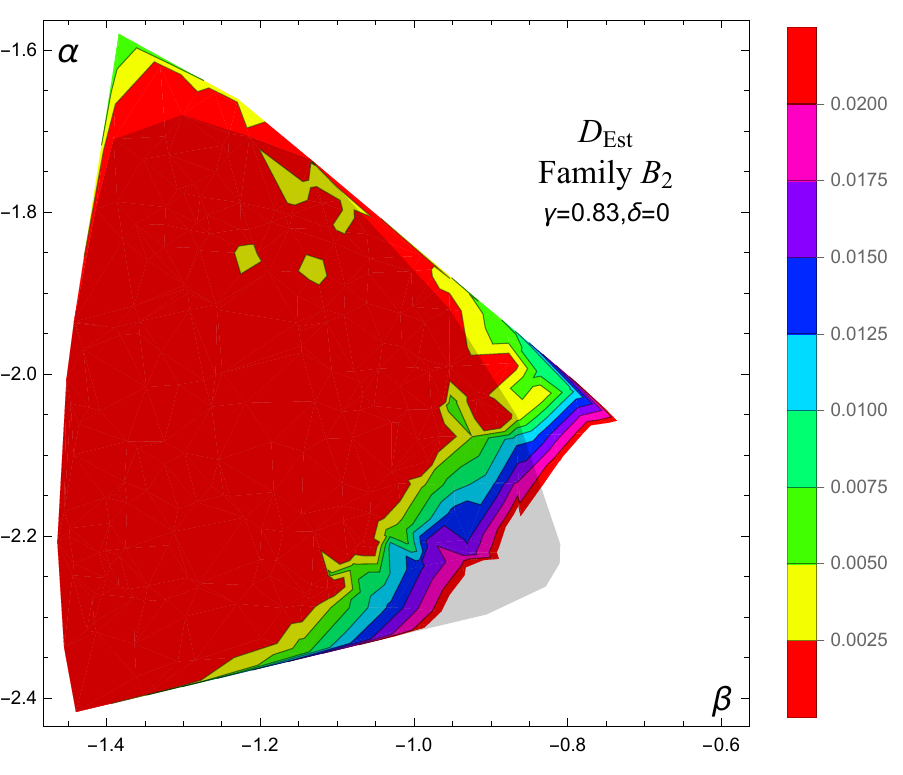}
    \end{subfigure}
    \begin{subfigure}[r]{0.3\textwidth}
    \includegraphics[width=60mm]{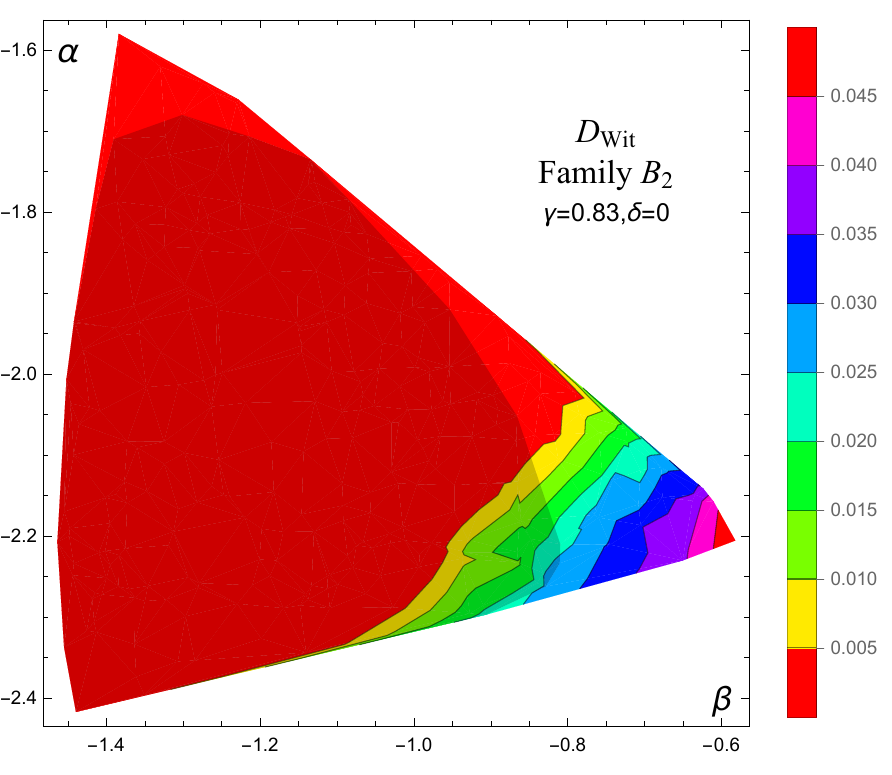}
    \end{subfigure}
    \caption{Interpolated plots of $D_{\text{Last}}$ (left), $D_{\text{Est}}$ (center), and $D_{\text{Wit}}$ (right) for PPT states of Family $B_2$ with up to 40000 corrections for selected states.}
\end{figure}
\end{widetext}\begin{widetext}
\begin{figure}[!h]
\centering
    \begin{subfigure}[l]{0.3\textwidth}
    \includegraphics[width=55mm]{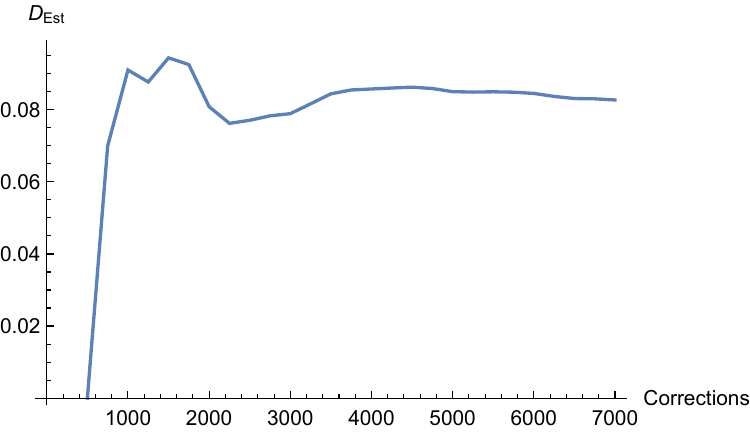}
    \end{subfigure}
    \begin{subfigure}[l]{0.3\textwidth}
    \includegraphics[width=55mm]{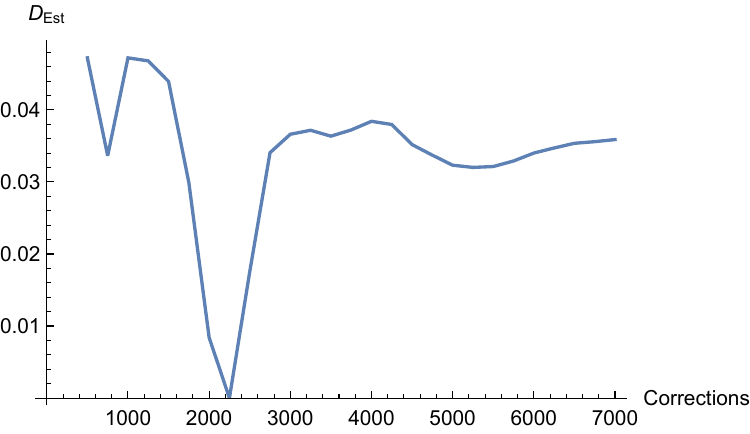}
    \end{subfigure}
    \begin{subfigure}[l]{0.3\textwidth}
    \includegraphics[width=55mm]{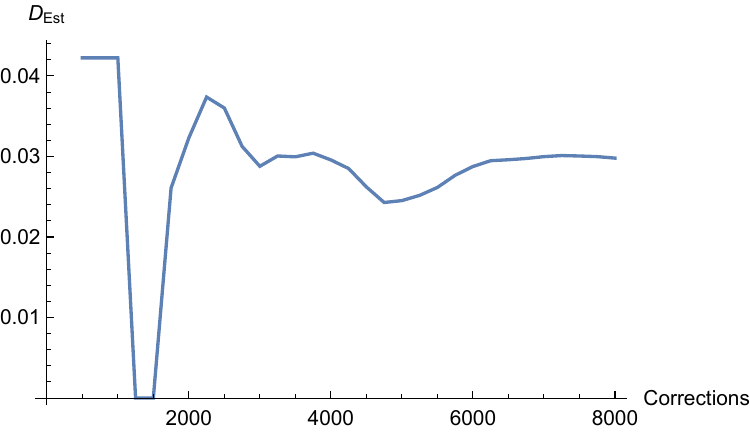}
    \end{subfigure}
    \vfill
    \begin{subfigure}[l]{0.3\textwidth}
    \includegraphics[width=55mm]{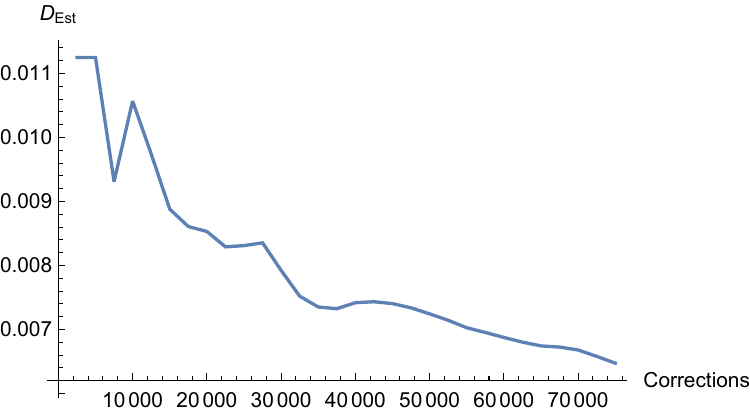}
    \end{subfigure}
    \begin{subfigure}[l]{0.3\textwidth}
    \includegraphics[width=55mm]{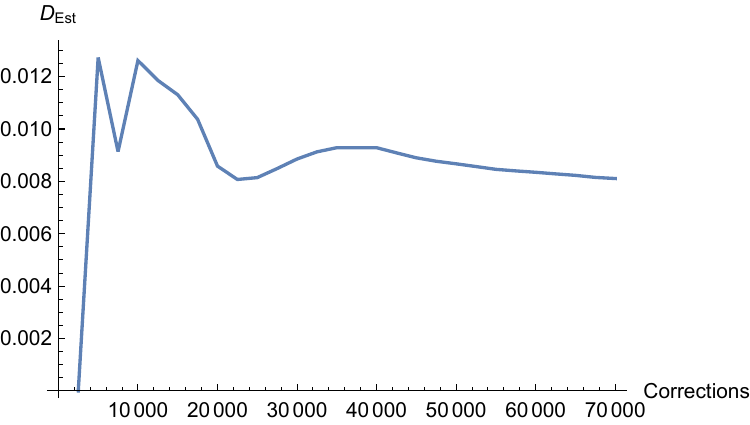}
    \end{subfigure}
    \begin{subfigure}[l]{0.3\textwidth}
    \includegraphics[width=55mm]{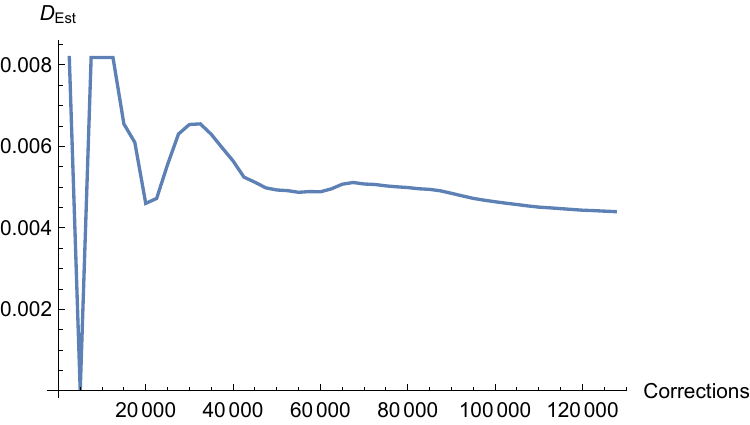}
    \end{subfigure}
    \caption{Dynamics $D_{\text{Est}}$ in function of the number of corrections for three arbitrarily selected states from Family $A$ (up) and $B$ (down).}
\end{figure}
\end{widetext}
Finally, let us stress low hardware requirements of the algorithm. A typical computation time  for a single state with up to 10000 corrections was around 20 minutes on an Intel Core i9-11900K processor (running sixteen single-core instances) and all instances together consumed about 1 GB of RAM at any time. It is thus both computationally and resource-wise efficient.

\section{Conclusions}
Until recently, even a rough classification of quantum states as separable or entangled has been a very demanding challenge, close to impossible. However, last years have brought in a significant progress. The Gilbert algorithm does not give a definite and final solution to this problem, but it can provide useful insight. First, it is able to provide incredibly precise approximations of some states, strongly hinting them as separable. On the other hand, it provides entanglement witness , directly certifying nonseparability. This leaves few ambiguous cases of very weakly entangled  states and those, for which subsequent corrections require particularly many trail states. This was the  case of Family $B_2$. In this contribution we have demonstrated that also a strictly statistical interpretation of the result utilizing $D_{\text{Est}}$ can give a reasonable approximation of a set of separable states. In at least one case we have certified entanglement where other combined analytical, numerical, and machine learning-based techniques have failed.

The presented technique of ``cartography of entanglement'' can be used universally applied to identify or estimate the boundary between separable and entangled states, regardless of the dimension, the number of subsystems, or a type of quantum correlations in question. For example, it can find a variety of applications in solid state models. Importantly, the algorithm has not been fed with any information other than the input state. It is irrelevant if the state has free or bound entanglement. The technique could be combined with machine learning and interpolation techniques, but it can also generate useful results. In contrast to FREE/SEP/BOUND categorization, it provides a qualitative information about entanglement. While new bound entangled states can be easily detected, it
is the question if their nonclassicality can be meaningful in an experimental realization.
\section{acknowledgements}
This work is a part of NCN
Grant  No. 2017/26/E/ST2/01008. MW acknowledges partial support by the Foundation for Polish
Science (IRAP project, ICTQT, Contract No. 2018/MAB/5,
co-financed by EU within Smart Growth Operational Programme).

\end{document}